\documentclass[aps,prd,preprint]{revtex4-1}
\usepackage{graphicx}

\def\be{\begin{equation}}	\def\ee#1{\label{#1}\end{equation}}
\def\ba{\begin{array}}	\def\ea{\end{array}}
\def\bea{\begin{eqnarray}}	\def\eea{\end{eqnarray}}
		\def\pa{\partial}
  \def\ci{\cite}

\def\Title#1{\begin{center} {\Large #1 } \end{center}}
\def\Author#1{\begin{center}{ \sc #1} \end{center}}
\def\Address#1{\begin{center}{ \it #1} \end{center}}

\newcommand\pubnumber{MAXLA-5/21, ``GenSol SE''}
\newcommand\pubdate{\today}
\newcommand\pubblock{\rightline{\begin{tabular}{l} \pubnumber\\
\pubdate \end{tabular}}}
\newenvironment{Abstract}{\begin{quotation}  }{\end{quotation}}

\newenvironment{Presented}
{\begin{quotation} 
\begin{center}\begin{large}}{\end{large}
\end{center} 
\end{quotation}}

\begin{document}

\begin{titlepage}
\pubblock

\vfill
\Title{ General solution of the Schr\"odinger equation} 
\Author{Mikhail N. Sergeenko}
\Address{ Institute of Radiobiology of the National Academy 
of Sciences of Belarus\\ BY-246007, Gomel, Belarus\\
{\rm msergeen@gmail.com}}
\vspace{10mm}

\centerline{\bf Abstract}
\begin{Abstract}
The wave equation in quantum mechanics and its general solution 
in the phase space are obtained. 

%
\end{Abstract}

\vspace{3mm}
\begin{Presented}
 Belarus --- 2021
\end{Presented}
\vspace{5mm}
\end{titlepage}

 The Schr\"odinger wave (SW) equation in quantum mechanics (QM) is 
usually solved in terms of special functions or numerically. 
 We use here the abbreviation SW which means ``Schr\"odinger's wave'' 
known as $\psi$-function. 
 The general approach to solve the SW equation is to reduce its 
to the equation for hypergeometric function or some special function. 
 To do that one needs to find first a special transformation for 
the wave function (w.f.) and its arguments to reduce the original 
equation to the hypergeometric form. 
 However, the solution of the SW equation can be obtained in 
elementary functions. 

 The static one-dimensional (1D) SW equation for a free particle 
of mass $m$~\ci{LandLif77,Bohm79,Shiff55},
\be 
\frac 1{2m}\left(-i\hbar\frac d{dx}\right)^2\psi(x)=E\psi(x),
\ee{ShrE}
has a general solution in the form of a superposition of two 
plane waves in configuration space, 
\be 
\psi(x) = C_1e^{ikx} + C_2e^{-ikx}. 
\ee{psix} 
 Here $C_1$ and $C_2$ are (in general) complex constants, 
$k=p/\hbar$ is the wave number defined as $k=2\pi/\lambda_B$ for 
the de Broglie wavelength $\lambda_B=h/p$ with the particle 
constant momentum $p=\sqrt{2mE}$, $h=2\pi\hbar$ is the Plank's 
constant. 

 There is another approach to solving the SW equation. 
 It is known that the SW equation can be derived with the help of 
the Bohr's correspondence principle~\ci{Bohm79}. 
 This fundamental principle has been used at the stage of creation 
of quantum theory. 
 It is used to establish correspondence between classical functions 
and operators of QM, and to derive the apparent form of 
the operators. 
 Moreover, the correspondence principle points out the way to 
a simplest solution of the SW equation.

 The correspondence principle states that the laws of quantum 
physics must be so chosen that in the classical limit, where many 
quanta are involved, the quantum laws lead to the classical 
equations as an average. 
 In this way, in~\ci{MyPRA96} this principle has been used to derive 
the non-relativistic quasi-classical (QC) wave equation appropriate 
in the QC region and the relativistic QC wave equation~\ci{MyMPLA97}. 

 We transform (\ref{ShrE}), which is a problem on eigenvalues of 
energy $E$, to the equivalent form 
\be 
\left(-i\hbar\frac d{dx}\right)^2\psi(x)=p_E^2\psi(x),
\ee{ShrP}
which is a problem on eigenvalues of $p_E^2=(\hbar k)^2=2mE$, and  
introduce the dimensionless phase variable $\phi=kx$.  
 This gives the SW equation in the phase space, 
\be 
\psi^{\prime\prime}_{\phi\phi} + \psi = 0, 
\ee{ShPhi}
which is the linear homogeneous second-order differential (LHD$_2$) 
equation in canonical form. 
 The general solution of this equation is given by a superposition 
of two plane waves in the phase space $\{\phi\}$, 
\be 
\psi(\phi) = C_1e^{i\phi} + C_2e^{-i\phi}.
\ee{Psifi} 
 The dimensionless phase variable $\phi=W/\hbar$ is written in 
terms of the reduced {\it classical} action, $W=px$, for a free 
particle. 
 The purpose of this letter is to show that the general solution 
of the SW equation for the case of an interacting particle has 
the same form (\ref{Psifi}) for the corresponding {\it reduced 
classical action} $W$. 

 In QM, the concept of ``action in QM'' is used. 
 In the semiclassical approximation, the ``action in QM'' is used 
in the form of an expansion in terms of the Plank's constant $\hbar$: 
$S=S_0+\hbar S_1+\dots$~\ci{From65}. 
 But, in the quantization condition the {\it classical} action 
$S_0$ is used, and the corrections $S_1$, $S_2,\dots$ have 
no physical meaning, but lead to artificial problems such as 
divergence of the solution at the turning points. 
 In our opinion, the concept of ``action in QM'' is NOT necessary 
for the theory. 
 Let us show that the SW equation and its solution can be obtained 
from the {\it classical} action. 

 In the general case of an interacting particle, we consider 
a conservative system when the Hamiltonian $H(x,\,p)=p^2/2m+V(x)$ 
is not an explicit function of time $t$. 
 The SW equation for the interaction potential $V(x)$, 
\be 
\left[\frac 1{2m}\left(-i\hbar\frac d{dx}\right)^2
+V(x)\right]\psi(x)=E\psi(x),
\ee{ShVx} 
can be written in the Sturm-Liouville canonical form
\be 
\psi^{\prime\prime}_{xx}+[k^2-U^2(x)]\psi=0,
\ee{PsiSL}
where $k^2=2mE/\hbar^2$, $U^2(x)=2mV(x)/\hbar^2$, 
$p^2(x)=k^2-U^2(x)$. 

 In this case the Hamilton's principal function (the generating 
function) is 
\be 
\psi(t,\,x)=C e^{i[-Et + W(x)]/\hbar},\quad W(x)=\int^x p(x)dx.
\ee{psWx} 
 The second derivative of this function with respect to 
the variable $x$ is
\be 
\psi^{\prime\prime}_{xx}=\left[\left(\frac i\hbar p\right)^2
+\frac i\hbar\frac{dp}{dx}\right]\psi,
\ee{ShVxIm} 
where $p(x)=(dW/dx)$. 
 Multiplying (\ref{ShVxIm}) by the constant $(\hbar/i)^2$, we get 
\be 
\left(\frac\hbar{i}\frac d{dx}\right)^2\psi
=\left[2m[E-V(x)]+\frac\hbar{i}\frac{dp}{dx}\right]\psi.
\ee{Shcx}
 The equality (\ref{Shcx}) gives the SW equation (\ref{ShVx}) if 
\be  
\frac\hbar{i}\frac{dp}{dx} = 0\quad
\Rightarrow\quad \bar{p}={\rm const},
\ee{dpdx0}

 What does this mean? 
 The operators in QM are Hermitian, i.e. their eigenvalues are real. 
 The operator $\hat p^2$ in the left hand side of (\ref{Shcx}) is 
Hermitian, if the condition (\ref{dpdx0}) is met. 
 The Plank's constant is {\it constant} value, not zero, hence,   
the derivative $p^\prime_x=0$. 
 Here in (\ref{dpdx0}) $\bar{p}=\bar{k}\hbar$ is momentum eigenvalue 
(constant value) and $\phi=\bar{k}x$ is the phase variable. 

 The condition (\ref{dpdx0}) is the key one in solving the SW 
equation and supplies the {\em Hermiticity} of the squared momentum 
operator, $\hat p^2$, in (\ref{Shcx}). 
 We emphasize that only if the condition (\ref{dpdx0}) is satisfied, 
it is possible to obtain from (\ref{Shcx}) the SW equation 
(\ref{ShVx}). 
 Functions in mathematics can take both continuous and discrete 
values for specific values of argument. 
 The requirement (\ref{dpdx0}) is a prerequisite for permitted 
movements in QM; it defines {\it allowed motions} or stationary 
states of a quantum system, i.e., the generalized momentum 
$p(x)$ can only take some constant discrete values (momentum 
eigenvalues $\bar{p}$). 
 Thus, equality (\ref{dpdx0}) leads from (\ref{Shcx}) to the SW 
Eq. (\ref{ShVx}) and its solution of the form~(\ref{psWx}). 

 Transform (\ref{PsiSL}) to the dimensionless phase variable 
$\phi(x)=W(x)/\hbar$. 
 This can be done using the identity 
\be 
\frac d{dx}\left(f\frac{dy}{dx}\right) 
=\left(\sqrt{f}\frac{d^2}{dx^2}-\frac{d^2}{dx^2}\sqrt{f}\right)
\left(\sqrt{f}y\right)
\ee{iden} 
that gives the equation
\be 
\Psi^{\prime\prime}_{\phi\phi} +[1-\delta(\phi)]\Psi=0, 
\ee{eqphs}
where $\Psi=\sqrt{p}\psi$. 
 The function (functional) 
\bea
\delta(\phi)=\frac 1{\sqrt{p}}\frac{d^2\sqrt{p}}{d\phi^2}
=\frac 12\frac{d\epsilon}{d\phi} + \frac 14\epsilon^2,
\label{deltF}\\  
\epsilon=\frac\hbar{p^2}\frac{dp}{dx}\label{epsi}
\eea
has the meaning of a potential in the phase space with 
the properties: 
1)\,$\delta(\phi)=0$ for $V(x)=0$, 2)\,$\delta(\phi)=\infty$ 
at the turning points (TPs) given by the Eq.  $p(x)=0$, 
3)\,$\delta(\phi)$ is a small quantity of higher order at 
other points~\ci{From65}. 
 We can call (\ref{eqphs}) the SW equation in the phase space. 
 The fulfillment of the inequality $\epsilon\ll 1$ is a necessary condition for the application of the QC approximation in QM. 

 The roots of the equation $p(x)=0$ separate the classically 
allowed region where $p(x)\ge 0$ from the classically forbidden 
region where $p(x)<0$. 
 The function $\delta(\phi)$ in (\ref{eqphs}) has the properties 
of the $\delta$-function: according to (\ref{dpdx0}) the quantity 
$\epsilon=k^\prime_x/k^2=p^\prime_x/p^2=0$, therefore, 
$\delta(\phi)=0$ excluding the TPs determined by the equality 
(\ref{dpdx0}); in the TPs $\delta(\phi)=\infty$. 
 The corresponding equations in these regions follow from 
(\ref{eqphs}) and are given by the system in the phase space 
$\{\phi$~\ci{MyClaSol03} 
\be
\left\{
\begin{array}{lc}
\Psi^{\prime\prime}_{\phi\phi} + \Psi=0,\quad p(x)>0,\\
\Psi^{\prime\prime}_{\phi\phi} - \Psi=0,\quad p(x)<0.
\end{array}
\right.
\ee{syst}
 The general solutions of these equations are 
\be
\left\{
\begin{array}{lc}
\Psi(\phi) = Ae^{i\phi} + Be^{-i\phi},\quad p(x)>0,\\
\Psi(\phi) = Ce^{\phi} + De^{-\phi},\,\quad\ p(x)<0.
\end{array}
\right.
\ee{gen2so}

 Solution in QM (w.f.) must be continuous and finite in 
the entire range ($-\infty,\,\infty$). 
 To build the physical solution in the entire range we need 
to merge the oscillating solution(s) in classically allowed 
region where $p(x)\ge 0$ with the exponentially decaying 
solution(s) in classically inaccessible regions where 
$p(x)<0$~\ci{MyPRA96}. 

 The functions (\ref{gen2so}) should smoothly merge into each 
other at the turning points. 
 Matching these functions and their first derivatives at 
the turning point $x_k$ gives two equalities 
\be
\left\{
\begin{array}{lc}
A + B = C + D,\\
iA - iB = -C + D,\\
\end{array}
\right.
\ee{syst}
which yields
\be
\left\{
\begin{array}{lc}
A = \left(Ce^{i\pi/4} + De^{-i\pi/4}\right)/\sqrt{2},\\
B = \left(Ce^{-i\pi/4}+ De^{i\pi/4}\right)/\sqrt{2}.\\
\end{array}
\right.
\ee{syspa}
 The connection formulas (\ref{syspa}) supply the continuous 
transition of the general solutions (\ref{gen2so}) into each 
other at the turning point $x_k$. 

 The most popular and important in applications are 
the two-turning point (2TP) problems~\ci{MyClaSol03,MyPRA96}. 
 For the 2TP problem, the entire interval ($-\infty,\infty$) is 
divided by the TPs $x_1$ and $x_2$ into three regions. 
 This leads with the help of the connection formulas (\ref{syspa}) 
to the quantization condition~\ci{MyClaSol03,MyPRA96}
\be 
\int_{x_1}^{x_2}\sqrt{2m[E-V(x)]}dx=\pi\hbar\left(n+\frac 12\right). 
\ee{qc2}
 The final solution for the 2TP problems (the state function) in 
the phase space is~\ci{MyClaSol03} 
\be
\Psi_n(\phi) = C_n\left\{
\begin{array}{lc} 
e^{\phi -\phi_1}, & x<x_1,\\
\sqrt{2}\cos(\phi -\phi_1 -\frac\pi 4), & x_1\le x\le x_2,\\
(-1)^ne^{-\phi +\phi_2}, & x>x_2,
\end{array}
\right.
\ee{Asol}
where $\phi(n,\,x)=k_nx$, 
$\phi_1=\phi(x_1)=-\pi(n+\frac 12)/2$, 
$\phi_2=\phi(x_2)=\pi(n+\frac 12)/2$~\ci{MyPRA96}. 
 Here we have took into account the fact that, for the stationary 
states, the phase-space variable $\phi(n,\,x)$ at the TPs $x_1$ 
and $x_2$ depends on quantum number $n$ and does not depend on 
the form of the potential. 
 The normalization coefficient,
\be
C_n=\sqrt{\frac{k_n}{\pi(n+\frac 12)+1}},
\ee{Cn}
is calculated from the normalization condition 
$\int_{-\infty}^\infty\left|\psi_n(x)\right|^2dx=1$.

 The solution (\ref{Asol}) describes {\it free motion} of 
a particle-wave in the enclosure (the enclosure being 
the interaction potential). 
 Therefore, in bound state region, the interaction of 
the particle-wave with the potential reduces to reflection of 
the wave by ``walls of the potential''. 
 The ``classical'' solution (\ref{Asol}) is general for all types 
of 2TP problems and allows to solve multi-turning point problems 
which represent a class of the ``insoluble'' (by standard methods) 
problems with more than two turning points~\ci{MyMPLA97}. 

 The oscillating part of (\ref{Asol}), 
\be 
\Psi_n(x) = \sqrt{\frac{2k_n}{\pi(n+\frac 12)+1}}
\cos\left(k_nx +\frac\pi 2 n\right),
\ee{fn0}
has the form of a {\it standing wave}. 
 The form of the phase variable $\phi(n,\,x)=k_nx+\pi n/2$ 
guaranties that the state functions $\Psi_n(\phi)$ are necessarily 
either symmetric ($n=0,2,4,\ldots$) or antisymmetric 
($n=1,3,5,\ldots$). 
 The function (\ref{fn0}) corresponds to the principal term of 
the asymptotic series in theory of the LHD$_2$ equations, which 
in QM gives the asymptote of the exact solution of the SW equation. 
 The quantization condition (\ref{qc2}) in our solution of the 1D SW 
equation (\ref{ShVx}) is not approximate. 
 It is exact, i.e. (\ref{qc2}) reproduces the exact energy spectra 
for {\it all} known solvable 2TP problems in 
QM~\ci{MyClaSol03,MyPRA96}. 

 The case of three-dimensional (3D) problems was studied in our 
works~\ci{MyPRA96,MyMPLA00}. 
 The derivation of the 3D wave equation (\ref{QC3}) can be 
performed siminarly to 1D case given above. 
 Consider the case of a central potential. 
 In spherical coordinates, the variables are separated and 
the generating function is 
\be
\psi(t,\,{\bf r}) = Ce^{-i[Et - W({\bf r})]/\hbar},
\ee{psS03}
where $W({\bf r})=W(r)+W(\theta)+W(\varphi)$. 
 The first derivative gives   
\be
\vec\nabla\psi(t,\,{\bf r})=
\frac i\hbar(\vec\nabla W)\psi(t,\,{\bf r}). 
\ee{derps3}
 The second derivative results in the SW equation, 
\be 
\left[\frac{(-i\hbar\vec\nabla)^2}{2m} + V(r)\right]\psi(\vec r) 
= E\psi(\vec r).
\ee{shr3}

 A QC analysis of (\ref{shr3}) was performed~\ci{MyMPLA00}. 
 The SW equation (\ref{shr3}) was reduced to the form of 
the classical HJ equation. 
 Separation of the resulting equation using the correspondence 
principle results in the three 1D equations in canonical 
form (\ref{PsiSL}), 
\bea 
\left[\hbar^2\frac{d^2}{dr^2} + 2m(E-V)
-\frac{\vec{M}^2}{r^2}\right]{\rm R}(r) = 0,
\label{RadEq}\\
\left[\hbar^2\frac{d^2}{d\theta^2} +
\vec M^2-\frac{M_z^2}{\sin^2\theta}\right]{\rm\Theta}(\theta)=0,
\label{AngEq}\\
\left[\hbar^2\frac{d^2}{d\varphi^2}+M_z^2\right]{\rm\Phi}(\varphi)=0,
\label{FiEq}
\eea
where $\vec M^2$, $M_z^2$ are the constants of separation and, at 
the same time, integrals of motion~\ci{MyPRA96}. 
 These equations are equivalent to the 3D equation 
\bea 
\left[(-i\hbar)^2\Delta^c + U(r)\right]\Psi(\vec r)
=p^2_E\Psi(\vec r),\label{QC3}\\
\Delta^c = 
\frac{\pa^2}{\pa r^2} + \frac 1{r^2}\frac{\pa^2}{\pa\theta^2}
+\frac 1{r^2\sin^2\theta}\frac{\pa^2}{\pa\varphi^2}\label{DelC}.
\eea
 Here (\ref{QC3}) is the wave equation in canonical form for 
the eigenvalue of the square of the momentum $p^2_E=2mE$; 
$\Delta^c$ is the canonical operator, $U(r)=2mV(r)$. 
 Note that (\ref{QC3}) is not the SW equation. 

 Solution of (\ref{QC3}) reproduces the exact energy spectra 
for {\it all} known solvable 2TP problems in 
QM~\ci{MyClaSol03,MyPRA96} and ``insoluble'' potentials with 
more than two turning points~\ci{MyMPLA97}. 
 More examples of application of our solution are in~\ci{MyEPJC12}. 

 Summarizing, we have shown that the generating function 
(\ref{psWx}), expressed in terms of the Hamilton's principal 
function, is the key object in the solution of the SW equation. 
 In our solution, there is no place for the ``action in quantum 
mechanics''. 
 This means we can start consideration with the classical action 
and continue by its quantization. 
 The general solution of the SW equation is given by the system 
(\ref{gen2so}). 

 In our solution method, we use the same technique for all types 
of problems. 
 The same simple rules and the general solution (\ref{gen2so})  
formulated for 2TP problems work for multi-TP problems, as well. 
 This approach can be easily generalized for the non-separable 
problems. 
 In this case the classical action can not be written as sum of 
actions for each variable (degree of freedom). 
 As a result, we cannot have a unique quantum number for each 
degree of freedom. 
 We will have one quantum number for the non-separable variables, 
and unique quantum number for each separable variable. 
 The quantization condition will be a multi-dimensional integral 
over the non-separable variables.
 In this sense, our approach can be considered as a general method 
for finding the solution of the SW equation. 

\bibliography{BiDaQM}

\end{document}